%% file: main.tex
\documentclass[sigconf, twocolumn]{acmart}
\renewcommand\footnotetextcopyrightpermission[1]{}
\usepackage{enumitem}
\usepackage{balance}
\usepackage{xspace}
\usepackage{soul}
\usepackage{multirow}
\usepackage{color}
\usepackage{amsmath}
\usepackage{subcaption}
\definecolor{bluekeywords}{rgb}{0.13,0.13,1}
\definecolor{greencomments}{rgb}{0,0.5,0}
\definecolor{redstrings}{rgb}{0.9,0,0}

\usepackage[capitalise]{cleveref}

\usepackage{booktabs}

\renewcommand{\paragraph}[1]{\vskip 0.05in \noindent {\bf #1.}}

\newcommand{\dataset}{\textit{CD}\xspace}   
\newcommand{\codedataset}{\textit{CCD}\xspace} 
\newcommand{\nlpdataset}{\textit{NLCD}\xspace} 
\newcommand{\codedatasettrain}{\textit{CCD-Train}\xspace}
\newcommand{\codedatasettest}{\textit{CCD-Test}\xspace}
\newcommand{\nlpdatasettrain}{\textit{NLCD-Train}\xspace}
\newcommand{\nlpdatasettest}{\textit{NLCD-Test}\xspace}

\begin{document}


\title{Evaluating AIGC Detectors on Code Content}


\author{Jian Wang}
\email{jian004@e.ntu.edu.sg}
\affiliation{
  \institution{Nanyang Technological University}
  \country{Singapore}
}
\author{Shangqing Liu}
\email{liu.shangqing@ntu.edu.sg}
\affiliation{
  \institution{Nanyang Technological University}
  \country{Singapore}
}
\author{Xiaofei Xie}
\email{xfxie@smu.edu.sg}
\affiliation{
  \institution{Singapore Management University}
  \country{Singapore}
}
\author{Yi Li}
\email{yi_li@ntu.edu.sg}
\affiliation{
  \institution{Nanyang Technological University}
  \country{Singapore}
}

\begin{abstract}
Artificial Intelligence Generated Content (AIGC) has garnered considerable attention for its
impressive performance, with ChatGPT emerging as a leading AIGC model that produces high-quality
responses across various applications, including software development and maintenance. Despite its
potential, the misuse of ChatGPT poses significant concerns, especially in education and
safety-critical domains. Numerous AIGC detectors have been developed and evaluated on natural
language data. However, their performance on code-related content generated by ChatGPT remains
unexplored.

To fill this gap, in this paper, we present the first empirical study on evaluating existing AIGC
detectors in the software domain. We created a comprehensive dataset including \textbf{492.5K}
samples comprising code-related content produced by ChatGPT, encompassing popular software
activities like Q\&A (\textbf{115K}), code summarization (\textbf{126K}), and code generation
(\textbf{226.5K}). We evaluated six AIGC detectors, including three commercial and three
open-source solutions, assessing their performance on this dataset. Additionally, we conducted a
human study to understand human detection capabilities and compare them with the existing AIGC
detectors. Our results indicate that AIGC detectors demonstrate lower performance on code-related
data compared to natural language data. Fine-tuning can enhance detector performance, especially
for content within the same domain; but generalization remains a challenge. The human evaluation
reveals that detection by humans is quite challenging.


\end{abstract}



\maketitle

\section{Introduction}
\input{Sections/intro}
\section{Background}
\input{Sections/background}

\section{Study Design}
\input{Sections/approach}

\section{Study Results}
\input{Sections/results}
\section{Threats to Validity}
\input{Sections/discussion}
\section{Related Work}
\input{Sections/related}

\section{Conclusion}
\input{Sections/conclusion}

\bibliographystyle{ACM-Reference-Format}
\bibliography{sample-base}

\end{document}

%% file: Sections/intro.tex
In recent years, Artificial Intelligence Generated Content (AIGC) has attracted significant attention and interest from academia and industry. AIGC refers to content that is generated by advanced generative AI techniques. With AI techniques becoming more advanced, the generated content shows significantly better quality and is being used in a wide range of tasks. ChatGPT~\cite{chatgpt}, released by OpenAI, has become one of the most attention-grabbing
approaches. ChatGPT is a large language model (LLM) fine-tuned from GPT-3.5 series with Reinforcement Learning from Human Feedback (RLHF)~\cite{christiano2017deep,
stiennon2020learning, wu2021recursively} to build a conversational AI system. The massive learnt
knowledge in GPT-3.5 series and the well-designed fine-tuning process from RLHF allow ChatGPT to generate high-quality responses to user questions in various domains and contexts.

ChatGPT has demonstrated remarkable proficiency in generating content across a diverse range of
domains. Its ability to comprehend context, adhere to instructions, and produce coherent contents,
makes it particularly well-suited for tasks such as drafting emails, generating articles, composing
poetry, crafting stories, and producing social media contents. Furthermore, OpenAI has
highlighted the capability of ChatGPT in software development~\cite{openaidemo}. ChatGPT has been widely used
in software development tasks such as writing documentation, creating user manuals, generating code
snippets, reviewing code and repairing code.

Although ChatGPT offers numerous benefits for users, it is important to consider the potential for abuse. In educational domain, for instance, there is a risk that students may use ChatGPT to cheat on exams or plagiarize assignments, which violates academic integrity.
To avoid abuses, some universities have restricted the use of ChatGPT, as shown in the recent report\footnote{\url{https://www.universityworldnews.com/post.php?story=20230222132357841}}.
Similarly, in the industry, the source of content generated by ChatGPT must be carefully
considered, especially in security and safety-critical scenarios. AI-generated content may have low
quality or contain errors (e.g., toxic content or bugs) that could lead to serious consequences
~\cite{bang2023multitask}.
For example, to prevent malicious use of the contents generated by ChatGPT when answering questions, Stack
Overflow has announced that the ChatGPT-generated content is temporarily
banned,\footnote{\url{https://meta.stackoverflow.com/questions/421831/temporary-policy-chatgpt-is-banned}}
because ``\emph{the average rate of getting correct answers from ChatGPT is too low,
the posting of answers created by ChatGPT is substantially harmful to the site and to users who are
asking and looking for correct answers}''.

With the increasing use of ChatGPT in a wide range of domains, including software development, it
becomes more crucial to develop effective tools to detect AI-generated contents. For example, many
AIGC detectors~\cite{Hello-SimpleAI-open,
GPT2-Detector-open, DetectGPT-open, Contentatscale-close, Copyleaks-close, GPTZero-close,
Sapling-close, Writefull-close, Writer-close, Openai-close, Compilatio-close} from both academia and
industry have been developed to detect the generated contents from GPT-series models, including GPT-2~\cite{radford2019language}, GPT-3~\cite{brown2020language}, and ChatGPT.
While these tools have been proposed to detect ChatGPT-generated content, it remains unclear how
effective these tools are, particularly in the context of the software development domains. 

To fill this gap, in this paper, we take an early step and conduct a comprehensive empirical study
to evaluate the existing detectors, including both the open-source and commercial ones, on their
capacities of detecting the code-related content (e.g., code and documents) generated by ChatGPT.
Specifically, the study aims to answer crucial questions as follows: \textit{How accurate are the
current tools for detecting code-related content generated by ChatGPT? What are the differences in
performance between detecting natural language content and code-related content generated by
ChatGPT? Can fine-tuning the detection tools enhance their capability to identify ChatGPT-generated
content? How robust are the detection tools in detecting content that has been modified based on
ChatGPT-generated content? How about the human detection capabilities compared with the existing
AIGC detectors?}









To conduct this study and answer these questions, we constructed two datasets, namely the Code-Related Content Dataset (\codedataset) and the Natural Language-Related Content Dataset (\nlpdataset), by generating related content using ChatGPT in the domains of programming and natural language, respectively.
\codedataset consists of \textbf{467.5K}
samples across three different code-related scenarios, i.e., Q\&A from stack overflow
(\textbf{115K}),
code-to-text generation~\cite{husain2019codesearchnet} (\textbf{126K}), and text-to-code
generation~\cite{iyer2018mapping, hendrycksapps2021} (\textbf{226.5K}).
\nlpdataset, which contains \textbf{25K} samples, was constructed by using ChatGPT to polish the content from Wikipedia~\cite{petroni-etal-2021-kilt}.
 Note that each sample in \codedataset and \nlpdataset is a pair including the human-generated data and ChatGPT-generated data.



Based on this dataset, we design comprehensive experiments to evaluate the capabilities of existing detection tools including three open-source detectors (GPT-2 Detector~\cite{GPT2-Detector-open},
RoBERTa-QA~\cite{Hello-SimpleAI-open} and DetectGPT~\cite{DetectGPT-open}) and three commercial
detectors (GPTZero~\cite{GPTZero-close}, Writer~\cite{Writer-close} and Text
Classifier~\cite{Openai-close}). We evaluate the performance of selected tools in detecting program contents generated by ChatGPT with those generated by human.
Additionally, we conducted a human study to understand human detection capabilities and compare them with the existing AIGC detectors.

Extensive experiments have revealed that current AIGC detectors struggle to detect code-related data compared to natural language data. Although fine-tuning is able to improve performance, however, the generalization capacities are limited. A human study also suggests that humans encounter similar difficulties, particularly when dealing with code data, which can be like blindly guessing due to its complexity. Overall, the main contributions of our paper are summarized as follows:
\begin{itemize}[leftmargin=*]
    \item We conducted a comprehensive empirical study to evaluate the performance of six AIGC detectors, including three open-source detectors and three commercial detectors, on detecting code-related content generated by ChatGPT. To the best of our knowledge, this is the first study that specifically evaluates the performance of different AIGC detectors on code-related content generated by ChatGPT.
    \item We construct two large-scale datasets namely \codedataset and \nlpdataset, consisting of
    \textbf{467.5K} code-related samples and \textbf{25K} natural language-related samples.
    \item We conduct a human study to study the difficulty of detecting content generated by ChatGPT and compare it to the performance of AIGC detectors.

\end{itemize}


%% file: Sections/background.tex
\begin{table*}[t]
\caption{The details of the existing AIGC detectors.}
\vspace{-4mm}
\label{tbl-dector-details1}
\small
\begin{tabular}{cl|cccc|cc|c}
  \toprule
               \multicolumn{2}{c|}{\multirow{2}{*}{Detector}}               &
               \multicolumn{4}{c|}{Supported
Models}       & \multicolumn{2}{c|}{Interfaces} & \multicolumn{1}{c}{\multirow{2}{*}{Input Length} } \\
                            \multicolumn{2}{c|}{}                           &   GPT-2    &
                            GPT-3    &  ChatGPT   &  Unknown   &  Website   &        API
                            &                                                     \\ 
                            \midrule
  \multirow{3}{*}{Open-source} & GPT2-Detector~\cite{GPT2-Detector-open}    & \checkmark
  &            &            &            &            &     \checkmark     &                $\leq$
  512 tokens                 \\
                               & DetectGPT~\cite{DetectGPT-open}            & \checkmark
                               &            &            &            & \checkmark &
                               \checkmark     &                  $\leq$ 512 tokens
                               \\
                               & RoBERTa-QA~\cite{Hello-SimpleAI-open}        &
                               &            & \checkmark &            & \checkmark &
                               \checkmark     &                  $\leq$ 512 tokens
                               \\ 
                               \midrule
  \multirow{8}{*}{Commercial}  & Contentatscale~\cite{Contentatscale-close} &            &
  \checkmark & \checkmark &            & \checkmark &                    &              25 words to
  25k chars               \\
                               & Copyleaks~\cite{Copyleaks-close}           &            &
                               \checkmark & \checkmark &            & \checkmark &
                               \checkmark     &               150 chars to 25k chars
                               \\
                               & GPTZero~\cite{GPTZero-close}               & \checkmark &
                               \checkmark & \checkmark &            & \checkmark &
                               \checkmark     &               250 chars to 20k chars
                               \\
                               & Sapling~\cite{Sapling-close}               &            &
                               \checkmark & \checkmark &            & \checkmark &
                               \checkmark     &                50 words to 20k chars
                               \\
                               & Writefull~\cite{Writefull-close}           &            &
                               \checkmark & \checkmark &            & \checkmark &
                               \checkmark     &                50 words to 2k words
                               \\
                               & Writer~\cite{Writer-close}                 &
                               &            &            & \checkmark & \checkmark &
                               \checkmark     &                  $\leq$1.5k chars
                               \\
                               & Compilatio~\cite{Compilatio-close}         &            &
                               \checkmark & \checkmark &            & \checkmark
                               &                    &                200 chars to 2k
                               chars                \\
                               & AITextClassifier~\cite{Openai-close}       &
                               &            & \checkmark &            & \checkmark
                               &                    &                   $\geq$1k
                               chars                    \\ 
                               \bottomrule
\end{tabular}
\end{table*}

\subsection{ChatGPT}
ChatGPT is a cutting-edge generative AI model developed by OpenAI, based on the GPT-3.5 architecture, which is the latest iteration in the GPT series of models, following 
GPT~\cite{radford2018improving}, GPT-2~\cite{radford2019language} and GPT-3~\cite{brown2020language}.
ChatGPT is primarily trained on large amounts of unlabeled text data. To enhance its ability to generate more natural and human-like responses in conversations, ChatGPT undergoes a supervised fine-tuning process using a combination of InstructGPT dataset and human conversations. During this process, the human conversations are transformed into a dialogue format and used in conjunction with the InstructGPT dataset to further train the model. 
ChatGPT utilizes reinforcement learning from human feedback (RLHF) algorithms to further refine its responses, which aligns a model trained on a general corpus of text data to that of complex human values. The combination of the massive knowledge learned from internet training data up to 2021 and the well-designed fine-tuning process enables ChatGPT to generate high-quality answers across a range of domains. Its strong conversational abilities have sparked interest in artificial general intelligence, and it has the potential to increase productivity in many industries. For example, it can be used as a writing assistant to help with tasks like summarizing, paraphrasing, and translating, or as a chatbot to hold conversations and answer questions on a variety of topics. 

However, as with any AI model, there are concerns about reliability~\cite{bang2023multitask},
ethics~\cite{zhuo2023exploring} and robustness~\cite{wang2023robustness}. Recent studies have explored potential issues that may arise with ChatGPT-generated content, such as cross-domain performance, ethical considerations in generating biased or sensitive content, and the robustness to adversarial attacks. These issues are largely due to ChatGPT's reliance on large amounts of training data. Given the potential issues associated with ChatGPT-generated content, it is important to use caution when relying on it, particularly in sensitive or specialized domains. To mitigate these risks, it is crucial to detect whether the content was generated by ChatGPT or other AI models to improve transparency and accountability. This will enable users to make informed decisions about the trustworthiness and reliability of the information presented, and reduce the potential for harm or misinformation.

\subsection{AIGC Detection}\label{sec:text-detection}
To ensure the responsible and ethical use of AI-generated content, various detectors have been developed to identify whether a given piece of content was generated by an AI model.  We have collected multiple detectors~\cite{Hello-SimpleAI-open,
GPT2-Detector-open, DetectGPT-open, Contentatscale-close, Copyleaks-close, GPTZero-close,
Sapling-close, Writefull-close, Writer-close, Openai-close, Compilatio-close} as up to March 2023, and their detailed information is presented in \cref{tbl-dector-details}.
The ``Supported Models'' column lists the types of models supported by the detectors, where
``Unknown'' means the supported model is unclear from the official documentation.
Column ``Interfaces'' indicates the supported interfaces which the detection tools may be accessed
from.
For example, ``Website'' denotes that the detector can only be accessed from its official website,
and ``API'' means it supports access from standard programming interfaces.
We can see that most commercial detectors support both GPT3 and ChatGPT detection while the
open-source detectors only support one type of detection (i.e., GPT2 or GPT3).
However, it is worth noting that some commercial detectors that support API access may not be free or may only allow a limited number of visits per day. 
For instance, Sapling~\cite{Sapling-close} provides API access at a cost of 25 dollars per
month and Writefull~\cite{Writefull-close} restricts accesses when the daily quota is reached.
Finally, as is shown in the ``Input Length'' column, each detector may have a different requirement
on the length of the input texts.
For example, the open-source detectors can only process input texts up to 512 tokens.
The commercial detectors accept input texts ranging from 25 words to 25K characters.

%% file: Sections/approach.tex
In this section, we give details on our study design.
Our study is centered around typical scenarios how ChatGPT are used to support software development
activities.
We collected data from both ChatGPT and human experts in each usage scenario and then compared the
performance of different detectors.
Next, we introduce the setup of each scenario, the detectors under comparison, and the research
questions to be studied.

\subsection{Scenarios and Data Collection}\label{sec:scenarios}
ChatGPT has been widely used in software development activities.
For example, it can be used to answer programming-related questions, summarize code snippets with
natural languages, and generate code based on natural language descriptions.
In this study, we focus on three of the most common scenarios in software development: (1) Q\&A on
programming topics, (2) code-to-text generation, and (3) text-to-code generation.
To conduct our study, we first collected relevant data from both human and ChatGPT. We then evaluated the capacity of different detectors in detecting ChatGPT-generated contents in these scenarios.
To better understand the performance of the detectors on code-related content, we also collected a new reference dataset consisting of purely natural language data (in the natural language polishing scenario) for comparison.
Next, we introduce the scenarios and process of data collection.


\subsubsection{Q\&A}\label{sec:qa}
It is a common practice for programmers to search the Internet for answers, when they have
questions on certain programming tasks.
Q\&A websites, such as Stack Overflow, are designed for this purpose.
Stack Overflow collects and organizes relevant answers, which become an essential resource for software developers today.
Stack Overflow expects high-quality answers from genuine experts to build a healthy and sustainable
community.
Due to concerns on the answer quality, posting answers generated by ChatGPT is banned on Stack
Overflow.
Yet, effective detection of AI-generated contents with high accuracy is necessary to enforce such a
policy.
Therefore, our first scenario focuses on studying the effectiveness of AIGC detectors in identifying programming-related answers generated by ChatGPT.

\paragraph{Data Collection}
To evaluate the effectiveness of AIGC detectors in the Q\&A scenario, we used the Stack Overflow dataset from Stack Exchange~\cite{stackexchange} which includes questions and answers posted on the Stack Overflow platform from September 2021 to November 2022.
For each question, we consider the top-voted response as the human-generated answer, and use ChatGPT to generate another answer in response to the same question.
In total, we obtained \textbf{115K} pairs of human-generated and ChatGPT-generated answers for the \textbf{115K} questions.
This dataset (denoted as Q\&A-GPT) provides a comprehensive benchmark for evaluating the
performance of detectors in identifying ChatGPT-generated content in the context of
programming-related questions and answers.

\begin{table*}[!t]
\vspace{-2mm}
\caption{The statistics of the collected data}
\vspace{-2mm}
\label{tbl-statistics-data}
\resizebox{1.5\columnwidth}{!}{


\begin{tabular}{c|ccc|ccc}
\hline 
Split & Wiki-GPT & Q\&A-GPT & Code2Doc-GPT & Doc2Code-GPT & CONCODE-GPT & APPS-GPT\tabularnewline
\hline 
\hline 
Train & 0 & 100K & 100K & 100K & 70K & 0\tabularnewline
Test & 25K & 15K & 26K & 26K & 23K & 7.5K\tabularnewline
\hline 
Total & 25K & 115K & 126K & 126K & 93K & 7.5K\tabularnewline
\hline 
\end{tabular}

}
\end{table*}
\subsubsection{Code-to-Text Generation}\label{sec:code2text}
Generating natural language descriptions of a given code snippet has been a long-standing research challenge widely studied in academia~\cite{ahmad2020transformer, liu2021retrievalaugmented, iyer2016summarizing}.
Accurate descriptions of code can help programmers better understand its functionality and improve software development efficiency.
However, writing accurate code comments is a time-consuming and laborious task.
ChatGPT has demonstrated excellent capabilities in generating natural language descriptions of code, making it a promising solution for automating this task.
Yet, it is still not advisable to replace human-written code comments with machine-generated code summaries, for the lack of quality guarantees.
Therefore, detectors that are capable of identifying code summaries generated by ChatGPT are necessary to discover massive use of machine-generated code comments.


\paragraph{Data Collection}
Specifically, we adopted the widely used benchmark CodeSearchNet~\cite{husain2019codesearchnet}, where each sample is a pair (\textit{code}, \textit{description}), across six programming languages including Ruby, Javascript, Go, Python, Java and PHP.
Each sample has a ``docstring'' field which contains the descriptions of the code produced by human experts.
To obtain code descriptions from ChatGPT, we designed a prompt that asks ChatGPT to generate the summary of the given code. The prompt is as follows:
\begin{quote}
\textit{You will be given a <LANG> function code and your task is to generate a detailed summary of its behavior and functionality. Your summary should clearly explain what the function does, how it works, and what input parameters and output values it expects. You should write your explanation in clear and concise language.
}
\\
\textit{
Code: <CODE>}
\end{quote}



where <LANG> is one of the six programming languages, and <CODE> is the target code we would like to summarize.
In total, we select \textbf{126K} samples from CodeSearchNet, and generate the answers using ChatGPT, denoted as Code2Doc-GPT.
Each programming language accounts for a different number of samples. The number of Ruby, JavaScript, Go, Python, Java, and PHP is 3,457, 7,974, 22,897, 34,639, 22,843, and 34,300 respectively.
This dataset provides a comprehensive benchmark for evaluating the performance of detectors in identifying natural language descriptions of code generated by ChatGPT across a range of programming languages.

\subsubsection{Text-to-code Generation}\label{sec:text2code}
With the advancements in AI technology, particularly the development of large language models, there has been a surge of interest in automatically generating code from natural language descriptions.
Recent works~\cite{openai2023gpt4, chen2022codet} have revealed that GPT-3 series are powerful at writing programs following
human instructions. For example, GPT-3.5 achieves an accuracy of 48.1\%
on Python coding tasks~\cite{chen2021evaluating}, the latest version GPT-4~\cite{openai2023gpt4} released by OpenAI even achieves an accuracy of 67.0\% on the same dataset.
The use of AI-generated code may be prohibited in some contexts, for example, due to policies on academic integrity.
Our third scenario studies the effectiveness of the existing detectors, to better understand the
technical feasibility of
distinguishing code generated by ChatGPT from those written by human.

\begin{figure*}[!t]
     \centering
         \begin{subfigure}[b]{0.49\linewidth}
         \includegraphics[width=\linewidth]{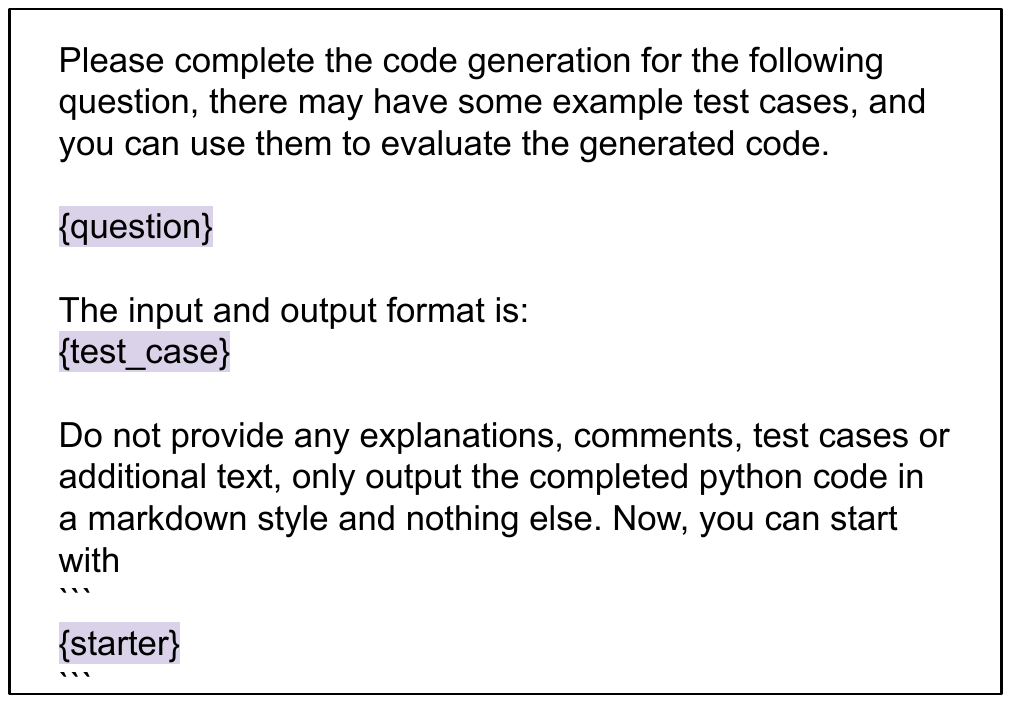}
         \caption{APPS prompt template}
         \label{fig:apps_prompt}
     \end{subfigure}
     \hfill
     \begin{subfigure}[b]{0.49\linewidth}
         \includegraphics[width=\linewidth]{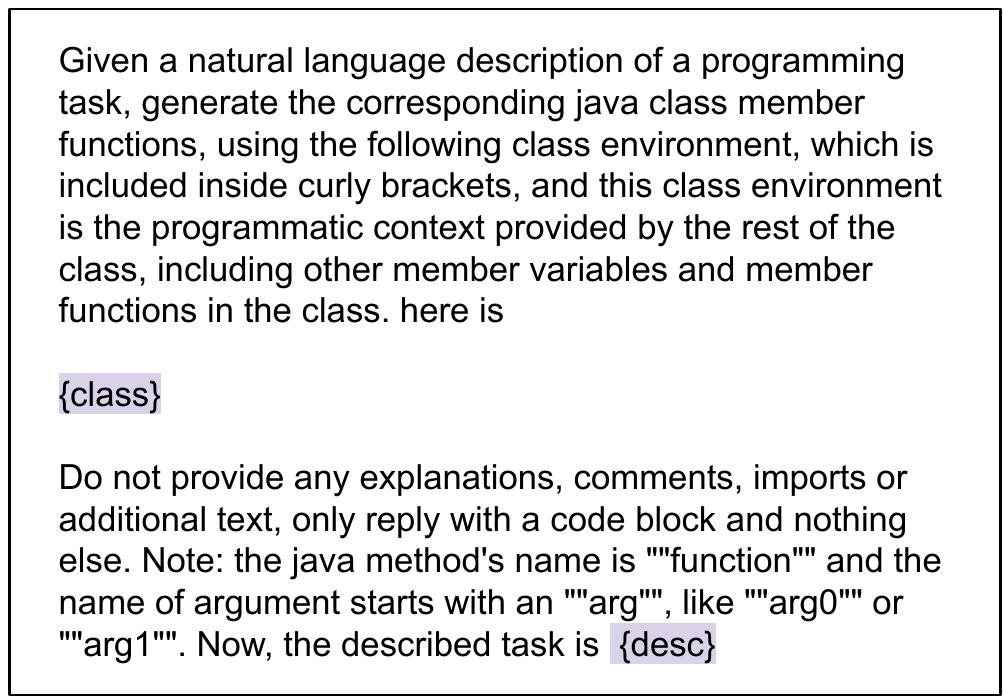}
         \caption{CONCODE prompt template}
         \label{fig:concode_prompt}
     \end{subfigure}
     \hfill
  \caption{The designed prompts for the dataset of APPs and CONCODE.}
\label{fig:prompts}
\end{figure*}

\paragraph{Data Collection}
Specifically, we collected the ChatGPT-generated code based on three different code
generation datasets, i.e., APPS~\cite{hendrycksapps2021}, CONCODE~\cite{iyer2018mapping}, and the
Code2Doc-GPT dataset described in \cref{sec:code2text}.
\begin{itemize}[leftmargin=*]
    \item The APPS dataset~\cite{hendrycksapps2021} is a Python dataset consisting of coding problems gathered from various public websites. Each problem in the dataset is accompanied by its description, ground-truth solutions, and test cases used to validate the implemented solutions. We regard the ground-truth solutions as the answers provided by human experts. To obtain solutions from ChatGPT, we created a prompt as shown in Figure~\ref{fig:apps_prompt}, where ``\{question\}'', ``\{test\_case\}'' and  ``\{starter\}'' are the placeholders of the problem description, the test cases and the function name provided in APPS.
    We finally collected \textbf{7.5K} samples from ChatGPT, denoted as APPS-GPT dataset.
    \item CONCODE~\cite{iyer2018mapping} is a Java dataset included in the CodeXGLUE~\cite{DBLP:journals/corr/abs-2102-04664} collection. Its goal is to generate class member functions for a Java class based on natural language descriptions and the programmatic context provided by the class environment, which includes member variables and other member functions in the class. To obtain answers from ChatGPT, we created a prompt as shown in Figure~\ref{fig:concode_prompt}, where  ``\{desc\}'' and ``\{class\}''  are the placeholders of the description and the class environment.
    The ground truth code provided in CONCODE is considered as the answer from human experts. We collected a total of \textbf{93K} samples, which we refer to as CONCODE-GPT dataset.
    \item The Code2Doc-GPT dataset constructed in \cref{sec:code2text} consists of detailed descriptions of given code generated by ChatGPT. We can naturally ask ChatGPT to generate code based on the code descriptions generated by ChatGPT in the Code2Doc-GPT dataset. To do so, we designed a prompt for generating code, as shown follows:

\begin{quote}
\textit{You will be provided with a detailed description of a <LANG> function, and your task is to
generate a <LANG> function that implements the program's behavior based on that description. You
should write the function code as accurately as possible based on the description, without
providing any additional explanations or assumptions. Your implementation should conform to the
standard of <LANG> syntax and coding conventions.},
\end{quote}

where <LANG> represents one of the programming languages, namely Ruby, Javascript, Go, Python, Java, or PHP. The original code in the Code2Doc-GPT dataset is considered to be the data from human experts. Additionally, we collected a total of \textbf{126k} code samples generated by ChatGPT, which we refer to as the Doc2Code-GPT dataset.
\end{itemize}
In the code generation scenario, we generated a total of \textbf{226.5K} pairs of code samples, where each pair consists of human-generated and ChatGPT-generated code.

\subsubsection{Natural Language Polishing}
To understand how the studied detectors may behave differently on code-related and natural-language contents, we also created a dataset consisting solely of natural language texts generated by ChatGPT.
This dataset serves as a reference for other code-related tasks.
The polishing task is to ask ChatGPT to rephrase a given paragraph of text with its own words.

\paragraph{Data Collection} We utilized ChatGPT to refine texts from the Wikipedia dataset~\cite{petroni-etal-2021-kilt}. To do so, we designed a prompt with the following format: ``\textit{Please polish the following content: <TEXT>}'', where <TEXT> is a placeholder for the raw text from the Wikipedia dataset. The original raw text is provided by human experts, while the polished version is the corresponding text generated by ChatGPT.
We collected a total of \textbf{25K} samples, which we refer to as the Wiki-GPT dataset.
In summary, we collect \textbf{492.5K} samples in total. The statistics of the \dataset are present in Table~\ref{tbl-statistics-data}. As one of our research questions is to explore the extent that fine-tuning can help on the collected dataset, we need to split the collected data into the train set and test set. Specifically, we randomly select 100K samples from the dataset of Q\&A-GPT, Code2Doc-GPT to construct the \nlpdatasettrain, 100K samples from Doc2Code-GPT and 70K samples from CONCODE-GPT to construct \codedatasettrain. The remaining data are used for the test set i.e., \nlpdatasettest and \codedatasettest.

\paragraph{Data preprocessing} Through our careful inspection of the generated ChatGPT answers, we found that some of them contained identifiable symbols or phrases indicating that they were generated by ChatGPT.
To improve the quality of our curated datasets, we removed all portions that may reveal the ground
truth.
Specifically, we removed sentences such as ``As an AI language model...'' or ``As a language model...'' from the generated contents if they appeared.
For code snippets, we only kept code blocks, while removing any extraneous contents surrounding them.


\subsection{Selected Detectors}
In this section, we introduce the AIGC detectors to be compared in this study. The detectors are divided into two categories: commercial detectors and open-source detectors. We select three following commercial detectors because large-scale testing can be conducted on these detectors. Some other commercial detectors may not be free or only allow a limited number of visits per day. In addition, to our knowledge, there are currently three open-source detectors and we take them for comparison.
\subsubsection{Commercial Detectors}
\begin{itemize}[leftmargin=*]
\item {GPTZero~\cite{GPTZero-close}.} According to the official website, it is a classification
model to predict whether the content is generated by a language model, providing predictions on a
sentence, paragraph and document level. Specifically, it is trained on a large corpus of
human-written and AI-generated text. The human-written text is from student-written articles, news
articles, as well as Q\&A datasets across multiple disciplines in the sciences and humanities. For
each human-written text, the AI-generated text is generated by the AI tool to ensure the training
dataset is balanced. The model is tested on a never-before-seen set of human and AI articles with
an accuracy of 99\% in correctly distinguishing human-written articles and 85\% accuracy in
correctly distinguishing AI-generated articles.
\item {Writer~\cite{Writer-close}.} It is a free tool to detect AI content. More technical details
have not been revealed on the official website. But there is a
study\footnote{\url{https://www.bloggersgoto.com/writer-com-ai-content-detector-review}} from the
Internet consisting of 10 human-written content and 10 AI-written content to evaluate its
effectiveness. The study results show that 4 out of 10 human-written content is correctly detected
and 8 out of 10 AI-written content is successfully detected. Hence, from the used 20 samples, it
seems that Writer is more accurate in detecting AI-written content than human-written content.
\item {AITextClassifier~\cite{Openai-close}.} It is an official detector released by OpenAI, which is a fine-tuned GPT model that predicts the probability of the text generated by AI. Specifically, the training dataset for this detector consists of AI-generated and human-written text where human-written text came from three sources including the Wikipedia dataset, the WebText dataset and a set of human demonstrations from InstrcutGPT~\cite{ouyang2022training} while AI-generated text is paired with the human-written text produced by the model. Then a detector is trained on this balanced dataset. An evaluation on the validation set, which follows the same distribution as the training set, and challenge set, which is composed of human-written completions and completions from a strong language model demonstrates it outperforms its previously published classifier GPT2-Detector~\cite{GPT2-Detector-open}.
\end{itemize}

\subsubsection{Open-source Detectors}
\begin{itemize}[leftmargin=*]
\item {GPT2-Detector~\cite{GPT2-Detector-open}.} It is a fine-tuned detector from
RoBERTa~\cite{liu2019roberta} released by OpenAI, where the training dataset is from the outputs of
the 1.5B-parameter GPT-2 model. From the official introduction on their
website,\footnote{\url{https://openai.com/research/gpt-2-1-5b-release}} the detection rates
achieve nearly 95\% for detecting 1.5B GPT-2 generated text.

\item {DetectGPT~\cite{DetectGPT-open}.} It is an open-source tool to detect the content generated
by GPT-2. Different from other detectors, which require training a separate classifier for
detection, it just compares the log probabilities of a candidate passage under the detected model
with the average log probability of several perturbations of the passage from another generic
pre-trained language model (e.g., BART~\cite{lewis2019bart}, T5~\cite{raffel2020exploring}). If the
perturbed passages have a lower average probability than the original by some margin, the candidate
passage is likely to be the model generated. The evaluation from in-distribution data and
out-of-distribution data demonstrates that DetectGPT is competitive with the learning-based
detectors GPT2-Detector~\cite{GPT2-Detector-open} on the in-distribution data while outperforming
it on the out-of-distribution data.

\item {RoBERTa-QA~\cite{Hello-SimpleAI-open}.} Guo et al. proposed to train a logistic regression
model on the GLTR Test-2 features~\cite{gehrmann2019gltr}, a RoBERTa-single model and a RoBERTa-QA
model on the Human ChatGPT Comparison Corpus dataset (HC3) dataset which consists of 37K questions
covering financial, medical, legal, psychological, and open domains to detect the content generated
by ChatGPT. Specifically, the logistic regression model leverages the token features from HC3
dataset for the classification. RoBERTa-based detectors fine-tune a pre-trained RoBERTa model where
RoBERTa-single model leverages the text generated by ChatGPT for fine-tuning while RoBERTa-QA model
leverages the text generated by ChatGPT combined with the input question for fine-tuning. The
evaluation confirms that the robustness of RoBERTa-based detector is better than the regression
model. Furthermore, RoBERTa-QA is better than RoBERTa-single in the detection.
In this paper, we select RoBERTa-QA model for the evaluation.
\end{itemize}


\subsection{Experimental Design and Research Questions}
In this section, we will present the designed research questions and the experimental setup for each research question.

\subsubsection{RQ1: How effective are existing detectors in detecting ChatGPT-generated content?}\label{sec:rq1}
We will evaluate the performance of six detectors on dataset \codedatasettest and \nlpdatasettest.

\noindent \textbf{Evaluation Setup:} Specifically, if the input length of an answer exceeds the largest sequence
length required for the selected detector, we truncate it to meet the requirement. For some detectors, a threshold
is required to distinguish ChatGPT data or human data by the specific detector. The output probability greater than the defined threshold indicates that the content is generated by ChatGPT. We follow the default settings of these detectors, where GPTZero was set to 0.8 and the remaining selectors were set to 0.5.


\subsubsection{RQ2: To what extent can fine-tuning improve detection performance?}
Since the detectors we compared were primarily designed for detecting natural language content, they may not perform optimally on our code-related dataset. Therefore, in this question, we aim to investigate whether fine-tuning can enhance the performance of the detectors.

\noindent \textbf{Experimental Setup:}
We selected the open-source detector RoBERTa-QA~\cite{Hello-SimpleAI-open} for fine-tuning. Specifically, we fine-tuned six detectors with the dataset \nlpdatasettrain and \codedatasettrain. The first three detectors fine-tuned with the training dataset of Q\&A-GPT, Code2Doc-GPT and the composite of Q\&A-GPT and Code2Doc-GPT, which are more related to detecting natural language data, while the other three detectors fine-tuned on CONCODE-GPT, Doc2Code-GPT and the composite of CONCODE-GPT and Doc2Code-GPT, which are more related to code data.  The fine-tuned detectors are evaluated on the same test dataset, i.e., \nlpdatasettest and \codedatasettest.

\begin{table*}[t]
\caption{Comparison results of six detectors, where bold number is the best performance in the corresponding column.}
\vspace{-2mm}
\label{tbl-tool-results}
\resizebox{2.1\columnwidth}{!}{

\begin{tabular}{l|ccccccccc|ccccccccc}
\hline
\multirow{3}{*}{Detectors} & \multicolumn{9}{c|}{NLCD-Test} & \multicolumn{9}{c}{CCD-Test}\tabularnewline
 & \multicolumn{3}{c}{Wiki-GPT} & \multicolumn{3}{c}{Q\&A-GPT} & \multicolumn{3}{c|}{Code2Doc-GPT} & \multicolumn{3}{c}{CONCODE-GPT} & \multicolumn{3}{c}{Doc2Code-GPT} & \multicolumn{3}{c}{APPS-GPT}\tabularnewline
 & AUC & FPR & FNR & AUC & FPR & FNR & AUC & FPR & FNR & AUC & FPR & FNR & AUC & FPR & FNR & AUC & FPR & FNR\tabularnewline
\hline
\hline
GPT2-Detector & 0.42 & 0.89 & \textbf{0.21} & 0.31 & 0.94 & 0.40 & 0.65 & 0.81 & 0.53 & \textbf{0.63} & 0.71 & \textbf{0.11} & 0.41 & 0.92 & \textbf{0.16} & 0.57 & 0.74 & 0.25\tabularnewline
DetectGPT  & 0.54 & 0.35 & 0.37 & 0.40 & 0.06 & 0.34 & 0.16 & 0.02 & 0.62 & 0.55 & \textbf{0.00} & 1.00 & 0.00 & \textbf{0.00} & 1.00 & 0.54 & \textbf{0.00} & 0.99\tabularnewline
RoBERTa-QA & 0.54 & 0.35 & 0.33 & 0.42 & 0.05 & 0.32 & 0.23 & 0.01 & 0.48 & 0.46 & \textbf{0.00} & 1.00 & 0.02 & \textbf{0.00} & 1.00 & 0.43 & \textbf{0.00} & 0.99\tabularnewline
\hline
GPTZero & \textbf{0.63} & \textbf{0.03} & 0.83 & 0.85 & \textbf{0.02} & 0.56 & 0.90 & \textbf{0.00} & 0.54 & 0.50 & 0.17 & 0.87 & \textbf{0.76} & 0.25 & 0.31 & 0.57 & 0.01 & 0.98\tabularnewline
Writer & 0.60 & 0.07 & 0.83 & 0.72 & 0.04 & 0.77 & 0.56 & 0.15 & 0.62 & 0.34 & 0.29 & 0.89 & 0.56 & 0.08 & 0.86 & 0.38 & 0.06 & 0.95\tabularnewline
AITextClassifier & 0.62 & 0.61 & 0.26 & \textbf{0.95} & 0.63 & \textbf{0.01} & \textbf{1.00} & 0.82 & \textbf{0.00} & 0.46 & 0.81 & 0.24 & 0.56 & 0.68 & 0.29 & \textbf{0.61} & 0.99 & \textbf{0.00}\tabularnewline
\hline
Avg & 0.56 & 0.38 & 0.47 & 0.61 & 0.29 & 0.40 & 0.58 & 0.30 & 0.46 & 0.49 & 0.33 & 0.69 & 0.39 & 0.32 & 0.60 & 0.52 & 0.30 & 0.69\tabularnewline
\hline
\end{tabular}
}
\end{table*}

\begin{table*}[]
\caption{Comparison results of different programming languages from Doc2Code-GPT.}
\vspace{-2mm}
\label{tbl-program-results}
\resizebox{2.0\columnwidth}{!}{

\begin{tabular}{l|ccc|ccc|ccc|ccc|ccc|ccc}
\hline
\multirow{2}{*}{Detectors} & \multicolumn{3}{c|}{Go} & \multicolumn{3}{c|}{Java} & \multicolumn{3}{c|}{Javascript} & \multicolumn{3}{c|}{PHP} & \multicolumn{3}{c|}{Python} & \multicolumn{3}{c}{Ruby}\tabularnewline
 & AUC & FPR & FNR & AUC & FPR & FNR & AUC & FPR & FNR & AUC & FPR & FNR & AUC & FPR & FNR & AUC & FPR & FNR\tabularnewline
\hline
\hline
GPT2-Detector & 0.38 & 0.96 & 0.10 & 0.40 & 0.92 & 0.18 & 0.38 & 0.89 & 0.23 & 0.39 & 0.90 & 0.22 & 0.46 & 0.91 & 0.12 & 0.46 & 0.89 & \textbf{0.10}\tabularnewline
DetectGPT  & 0.00 & \textbf{0.00} & 1.00 & 0.00 & \textbf{0.00} & 1.00 & 0.00 & \textbf{0.00} & 1.00 & 0.00 & \textbf{0.00} & 1.00 & 0.00 & \textbf{0.00} & 1.00 & 0.00 & \textbf{0.00} & 1.00\tabularnewline
RoBERTa-QA & 0.01 & \textbf{0.00} & 1.00 & 0.02 & \textbf{0.00} & 1.00 & 0.01 & \textbf{0.00} & 1.00 & 0.02 & \textbf{0.00} & 1.00 & 0.00 & \textbf{0.00} & 1.00 & 0.05 & \textbf{0.00} & 1.00\tabularnewline
\hline
GPTZero & 0.59 & 0.02 & 1.00 & \textbf{0.90} & 0.13 & \textbf{0.09} & \textbf{0.90} & 0.09 & \textbf{0.15} & 0.55 & 0.68 & \textbf{0.21} & \textbf{0.92} & 0.08 & \textbf{0.11} & \textbf{0.75} & 0.06 & 0.59\tabularnewline
Writer & 0.52 & 0.03 & 0.96 & 0.60 & 0.08 & 0.84 & 0.63 & 0.11 & 0.79 & \textbf{0.61} & 0.10 & 0.80 & 0.51 & 0.09 & 0.89 & 0.52 & 0.11 & 0.91\tabularnewline
AITextClassifier & \textbf{0.95} & 0.60 & \textbf{0.01} & 0.47 & 0.51 & 0.52 & 0.37 & 0.87 & 0.26 & 0.52 & 0.78 & \textbf{0.21} & 0.43 & 0.74 & 0.38 & 0.44 & 0.35 & 0.74\tabularnewline
\hline
Avg & 0.41 & 0.27 & 0.68 & 0.40 & 0.27 & 0.60 & 0.38 & 0.33 & 0.57 & 0.35 & 0.41 & 0.57 & 0.39 & 0.30 & 0.58 & 0.37 & 0.24 & 0.72\tabularnewline
\hline
\end{tabular}
}
\end{table*}

\subsubsection{RQ3: How robust are these detectors when the ChatGPT-generated data is slightly modified?} In real-world scenarios, the content generated by ChatGPT may not be used directly, and some of the generated content can be modified for customization or to avoid detection. Therefore, we aimed to investigate the robustness of detectors, specifically whether the AIGC content could be detected when modified by different mutations.

\noindent \textbf{Experimental Setup:} We selected three mutation operations to modify the text data and three mutation operations to modify the code data. We applied these mutators to both the human-generated and ChatGPT-generated data, and evaluated the performance of the detectors on the mutated dataset. For each test data, we applied each of the three mutation operations (where applicable) to generate the mutated data.

The three mutations for the code-based data are as follows:
\begin{itemize}[leftmargin=*]
    \item \textit{Function Name Renaming}. The function name is replaced with a new name randomly selected from a function name list, which is constructed from CodeSearchNet~\cite{husain2019codesearchnet}.
    \item  \textit{Variable Name Renaming}. We randomly select one variable in the function and replace the name with a new name randomly selected from a variable name list, which is constructed from CodeSearchNet.
    \item  \textit{Code Statement Insertion}. We insert a new assignment statement where the variable name is changed with a new name for statement insertion. The new variable name is randomly selected from a variable name list, constructed from CodeSearchNet.
\end{itemize}

We used the existing NLP mutations~\cite{ma2019nlpaug} to modify the text data:
\begin{itemize}[leftmargin=*]
    \item \textit{Character Insertion}. Randomly insert a character from a word of a text sample.
    \item \textit{Character Deletion}. Randomly delete a character from a word of a text sample.
    \item \textit{Character Replacement}. Randomly replace a character with another random character in a word of a text sample.
\end{itemize}



\subsubsection{RQ4: How well can human distinguish contents generated by ChatGPT?} To answer this research question, we conducted an online
survey with experienced software developers and observed their ability in distinguishing machine-generated contents.

\paragraph{Experiment Setup}
In our study, we invited \textbf{50} experienced developers who have at least five years of programming experience to participate in an online survey.
By March 2023, \textbf{27} of them have provided valid responses.
We randomly divided the participants into two groups, namely, the ``Example'' and ``No-example'' groups, with 14 and 13 responses, respectively.
The participants from the ``Example'' group were shown an example for each task, which consists of a pair of contents generated by humans and ChatGPT, with the ground truth clearly labeled.
The ``No-example'' group was not shown any example.

The questionnaire consists of 50 questions covering 5 types of code-related data (i.e., Q\&A-GPT, Code2Doc-GPT, APPS-GPT, CONCODE-GPT, Doc2Code-GPT), with 10 questions per type.
For each question, either a natural-language text block or a code snippet is shown to the participants, and they may choose from one of the three options: ``Human'', ``ChatGPT'', and ``Unclear''.
The participants were given at most 100 seconds for each question, and at the end of the questionnaire, they were asked to indicate the most important factors that influenced their decisions.

\subsubsection{Evaluation Metrics}
To evaluate the performance of our detectors, we used their AUC scores as the evaluation metric, which are commonly used in existing works such as AITextClassifier~\cite{Openai-close}. We further add FPR and FNR as the evaluation metrics.

\noindent \textbf{AUC score.} The AUC score of a detector is interpreted as the probability that the model's ability to
accurately classify classes on a scale from 0 to 1, where 1 is best and 0.5 is as good as a random choice.
For example, an AUC score of 0.5 implies that the model is only as good as the random choice when assigning probabilities to samples.
The higher the AUC score of a classifier, the better its ability to distinguish between positive and negative classes.
We refer to the data generated by ChatGPT as the positive class.

\noindent \textbf{FPR.} It refers to the false positive rate, calculated as $FPR = \frac{FP}{FP + TN}$ where $FP$ is the number of false positives (i.e., samples incorrectly classified as ChatGPT-generated), $TN$ is the number of true negatives (i.e., samples correctly classified as human experts) and $N = FP + TN$ is the total number of ground truth negatives (i.e., samples labeled as human experts).

\noindent \textbf{FNR.} It refers to the false negative rate, calculated as $FNR = \frac{FP}{FN + TP}$ where $FN$ is the number of false negatives (i.e., samples incorrectly classified as human experts), $TP$ is the number of true positives (i.e., samples correctly classified as ChatGPT-generated) and $ P = FN + TP$ is the total number of ground truth positives (i.e., samples are labeled as ChatGPT-generated).

A lower FPR indicates that the model is less likely to mislabel human contents as machine generated, while a lower FNR indicates the reverse.


%% file: Sections/results.tex
In this section, we present the experimental results along with our analysis in an attempt to answer each research question.

\subsection{RQ1: Effectiveness of the Existing Detectors}
The performance of the six detectors on different datasets is shown in \cref{tbl-tool-results}.
The first column lists the detectors under comparison, the second column shows the results on \nlpdatasettest, and the third column shows the results on \codedatasettest. The AVG row displays the average value of each metric on the specific test set across all six detectors.

In general, the results show that detecting ChatGPT-generated content is challenging, with average AUC, FPR, and FNR on \nlpdatasettest at 0.58, 0.32, and 0.44, respectively, and on \codedatasettest at 0.46, 0.32, and 0.66, respectively.
The overall results show that detecting ChatGPT-generated contents is still very challenging.
Specifically, the AUC value is higher on \nlpdatasettest than on \codedatasettest, while the FNR is lower on \nlpdatasettest than on \codedatasettest. These results suggest that detecting ChatGPT-generated code is even more difficult than detecting natural language contents, potentially due to the fact that existing detectors are trained with more natural language data than code data.



\begin{table}[t]
\caption{Results of fine-tuned models (based on RoBERTa-QA) on different \nlpdatasettrain datasets.}
\label{tbl-fine-tuning-nl}
\vspace{-4mm}
\resizebox{1.0\columnwidth}{!}{
\begin{tabular}{l|ccc|ccc|ccc}
\hline
\multirow{3}{*}{NLCD-Train} & \multicolumn{9}{c}{NLCD-Test}\tabularnewline
 & \multicolumn{3}{c|}{Wiki-GPT} & \multicolumn{3}{c|}{Q\&A-GPT} & \multicolumn{3}{c}{Code2Doc-GPT}\tabularnewline
 & AUC & FPR & FNR & AUC & FPR & FNR & AUC & FPR & FNR\tabularnewline
\hline
\hline
RoBERTa-QA & 0.54 & \textbf{0.35} & 0.33 & 0.42 & 0.05 & 0.32 & 0.23 & 0.01 & 0.48\tabularnewline
\hline
Q\&A-GPT & 0.76 & 0.53 & 0.16 & \textbf{0.99} & 0.02 & 0.07 & \textbf{1.00} & \textbf{0.00} & 0.03\tabularnewline
Code2Doc-GPT & 0.66 & 0.51 & 0.26 & 0.93 & \textbf{0.01} & 0.54 & \textbf{1.00} & \textbf{0.00} & \textbf{0.00}\tabularnewline
Composite-NL & \textbf{0.78} & 0.63 & \textbf{0.10} & \textbf{0.99} & 0.06 & \textbf{0.02} & \textbf{1.00} & \textbf{0.00} & \textbf{0.00}\tabularnewline
\hline
\end{tabular}

}
\end{table}
By analyzing the results of the false positive rate (FPR) and false negative rate (FNR) of existing detectors, we can find that they often exhibit bias towards predicting positive or negative cases. Specifically, some detectors have higher FPR but lower FNR, indicating a tendency to predict positive results (i.e., generated by ChatGPT), while others have higher FNR but lower FPR, indicating a tendency to predict negative results (i.e., generated by humans). For example, RoBERTa-QA has a FPR and FNR of 0 and 1, respectively, indicating that all samples are predicted as negative (i.e., generated by humans).
We conjecture that the bias may be due to differences in the training data used by these detectors and the threshold settings for distinguishing positive and negative cases.
For example, although GPT2-Detector and RoBERTa-QA both use RoBERTa to train classifiers, GPT2-Detector uses the output of GPT-2 for training, while RoBERTa-QA uses the output of ChatGPT. Additionally, the thresholds used in the detectors may not be optimal for our dataset. For example, the AUC of AITextClassifier on Code2Doc-GPT is 1, but its FPR is 0.82. We selected the default thresholds in these detectors, but it may not be the best one for our dataset. Hence, selecting a proper threshold for different testset can be another challenge in AIGC detection.

Overall, commercial AIGC detectors tend to outperform open-source detectors in detecting
ChatGPT-generated content. Specifically, we observe higher AUC scores for commercial detectors on
the \nlpdatasettest. Among the evaluated detectors, AITextClassifier shows better performance on
the \nlpdatasettest. In addition, we also observe that the AUC scores of the three commercial
detectors on the Wiki-GPT dataset are lower compared to Q\&A-GPT and Code2Doc-GPT datasets, which
could be attributed to the high text similarity between the Wiki-GPT data generated by ChatGPT and
the original Wikipedia content,  as the generated Wiki-GPT data is polished from Wikipedia content.

In the \codedatasettest, we observed that none of the tested AIGC detectors were able to achieve satisfactory performance. GPTZero and AITextClassifier demonstrated promising results on Doc2Code-GPT and APPS-GPT (e.g., AUC scores of 0.76 and 0.61, respectively), but they were still not very effective. Among the three code-related datasets, we found that all three commercial detectors performed worse on CONCODE-GPT. Further analysis revealed that the size of functions in CONCODE-GPT was quite small, making the generated data from ChatGPT and human quite similar.

We further study the performance of detectors on code with different programming languages. The comparative detection results on the Doc2Code-GPT dataset (including 6 programming languages) are presented in Table~\ref{tbl-program-results}. Our analysis reveals that the detectors exhibit different performance levels in different programming languages. For instance, GPTZero performs remarkably well in detecting Java, JavaScript, Python, and Ruby code with AUC scores of 0.90, 0.90, 0.92, and 0.75, respectively, while it does not perform well on Go and PHP (only 0.59 and 0.55). Conversely, AITextClassifier shows excellent results in detecting Go language code (AUC score 0.95) but performs poorly on the remaining languages. It is worth noting that while GPTZero performs well on Java code in the Doc2Code-GPT dataset, it performs poorly on the CONCODE-GPT dataset (with an AUC score of 0.50), which is also written in Java. This indicates that the performance of the detector is sensitive to different datasets with varying distributions, even if they are written in the same programming language.

 \vspace{5pt}\noindent \fbox{
	\parbox{0.95\linewidth}{\textbf{Answers to RQ1}:
 Existing AIGC detectors generally perform better on natural language data than on code data, indicating that detecting ChatGPT-generated code is a more challenging task. Although commercial detectors outperform open-source ones, they still face difficulties in detecting ChatGPT-generated code.
 }
}

\begin{table}[t]
\caption{Results of fine-tuned models (based on RoBERTa-QA) on different \codedatasettrain datasets.}
\label{tbl-fine-tuning-code}
\vspace{-4mm}
\resizebox{1.0\columnwidth}{!}{
\begin{tabular}{l|ccc|ccc|ccc}
\hline
\multirow{3}{*}{CCD-Train} & \multicolumn{9}{c}{CCD-Test}\tabularnewline
 & \multicolumn{3}{c|}{CONCODE-GPT} & \multicolumn{3}{c|}{Doc2Code-GPT} & \multicolumn{3}{c}{APPS-GPT}\tabularnewline
 & AUC & FPR & FNR & AUC & FPR & FNR & AUC & FPR & FNR\tabularnewline
\hline
\hline
RoBERTa-QA & 0.46 & \textbf{0.00} & 1.00 & 0.02 & \textbf{0.00} & 1.00 & 0.43 & \textbf{0.00} & 0.99\tabularnewline
\hline
CONCODE-GPT & 0.95 & 0.11 & 0.12 & 0.82 & 0.65 & \textbf{0.02} & 0.73 & 0.68 & \textbf{0.06}\tabularnewline
Doc2Code-GPT & 0.90 & 0.37 & 0.10 & 0.93 & 0.39 & 0.04 & 0.76 & 0.61 & 0.08\tabularnewline
Composite-Code  &  \textbf{0.98} & 0.12  & \textbf{0.05} &  \textbf{0.94}  & 0.35  & 0.04  & \textbf{0.77} & 0.60  & 0.08 \tabularnewline
\hline
\end{tabular}

}
\end{table}




\subsection{RQ2: Performance of Fine-tuning}\label{sec:fine-tuing-performance}
Considering that existing detectors primarily focus on natural language content, we further investigate whether fine-tuning on code datasets can enhance detection performance for code data. We fine-tuned RoBERTa-QA with different training samples.
Table~\ref{tbl-fine-tuning-nl} presents the results of fine-tuning RoBERTa-QA using various training datasets from \nlpdatasettrain. We then evaluate this model on different test sets from \nlpdatasettest.
The second row displays the performance of the original detector as a reference for comparison with the three newly fine-tuned detectors. The "Composite" row refers to combining both training sets (i.e., Q\&A-GPT and Code2Doc-GPT from \nlpdatasettrain) to fine-tune the model. Similarly, Table~\ref{tbl-fine-tuning-code} presents the results of fine-tuning using training datasets from \codedatasettrain. Note that we did not use the APPS-GPT training set for fine-tuning, as it contains only a limited number of samples (75k in total).

The results show that fine-tuning can significantly enhance the performance of the existing detectors, both for natural language and code data. The fine-tuned models achieved high AUC scores ranging from 0.9 to 1.0 and low FPR/FNR, with most being less than 0.1, on the corresponding test dataset, except for Wiki-GPT and APPS-GPT. The poor results on Wiki-GPT and APPS-GPT are due to the lack of fine-tuning with the Wiki-GPT and APPS-GPT dataset. The overall results highlight the significance of fine-tuning in improving the performance of detectors in various domains

Furthermore, we observed that the fine-tuned models trained on one dataset have certain generalization abilities to other datasets. For instance, a fine-tuned model trained on Q\&A-GPT performs better on Code2Doc-GPT compared to its original detector. The AUC of this model improves from 0.23 to 1.00 in Code2Doc-GPT, while FPR and FNR decrease to 0.00 and 0.03, respectively. Similarly, when a model is fine-tuned on Doc2Code-GPT, the AUC improves from 0.46 to 0.90 in CONCODE-GPT and from 0.43 to 0.76 in APPS-GPT. It is worth mentioning that even though the fine-tuned models are not trained on Wiki-GPT and APPS-GPT, they still achieve significant improvement. Hence, we conjecture that some common patterns across different scenarios may appear in the ChatGPT-generated content, and the fine-tuned models can learn them.
Finally, we find that by combining the training data (i.e., Composite-NL in Table~\ref{tbl-fine-tuning-nl} and Composite-Code in Table~\ref{tbl-fine-tuning-code}), the fine-tuned model achieve further improvement.

 \vspace{5pt}\noindent \fbox{
	\parbox{0.95\linewidth}{\textbf{Answers to RQ2}: Fine-tuning on the collected ChatGPT-generated content can significantly improve the detection performance of the detectors. Furthermore, the fine-tuned models trained on one dataset have certain generalization abilities to other datasets.}
}

\subsection{RQ3: Robustness Analysis}
In this section, we evaluate the robustness of the fine-tuned models introduced in RQ2, particularly when the ChatGPT-generated data is modified. The results are shown in Table~\ref{tbl-robustness-results}, where the first column represents the two fine-tuned models based on the composite datasets in Table~\ref{tbl-fine-tuning-nl} and Table~\ref{tbl-fine-tuning-code}, respectively. The second column shows the six test datasets that were mutated using our mutation operators. The following columns present the AUC, FPR, and FNR scores. The column $C_{con}$ and $H_{con}$ show the ratio of samples that can be correctly detected before and after mutation, representing the consistency of the prediction.


In comparing the results of Table~\ref{tbl-robustness-results} with the results in the last rows of Table~\ref{tbl-fine-tuning-nl} and Table~\ref{tbl-fine-tuning-code}, we can observe a slight decrease in performance overall, but not to a significant extent in both natural language and code data. This could be due to our conservative mutation operators, which only change a small part of the content, and thus may not affect the model too much. However, there is a notable exception, where the performance drops significantly on the mutation version of Wiki-GPT, indicating low robustness on this dataset. This could be attributed to the fact that we did not include the Wiki-GPT data in the fine-tuning training dataset. We will try more mutation strategies in future.

By analyzing the results in columns $C_{con}$ and $H_{con}$, we can observe that, in natural language data, the models still perform well on human data but make some errors in ChatGPT-generated data that were correctly predicted before are now missed. On the contrary, in code data, the models can still detect the modified code correctly, but some human code that was correctly predicted before now cannot be detected. However, considering the overall performance of the models is not decreased a lot, there are cases where data that was previously mispredicted is now correctly predicted. 



 \vspace{5pt}\noindent \fbox{
	\parbox{0.95\linewidth}{\textbf{Answers to RQ3}:
The fine-tuned models are generally robust under our conservative mutation operators. However, the robustness of the models can be affected in datasets that were not included in the fine-tuning dataset.
 }
}

\begin{table}[t]
\caption{Robustness analysis of the fine-tuned models.}
\label{tbl-robustness-results}
\small
\resizebox{1.0\columnwidth}{!}{
\begin{tabular}{c|c|ccc|cc}
\hline 
Model & Mutation Set & AUC & FPR & FNR & $C_{con}$ & $H_{con}$\tabularnewline
\hline 
\hline 
\multirow{3}{*}{Composite-NL} & Wiki-GPT & 0.68 & 0.33 & 0.39 & 0.39 & 0.99\tabularnewline
 & QA-GPT & 0.98 & 0.02 & 0.15 & 0.73 & \textbf{1.00}\tabularnewline
 & Code2Doc-GPT & \textbf{1.00} & \textbf{0.00} & \textbf{0.01} & 0.98 & \textbf{1.00}\tabularnewline
\hline 
\multirow{3}{*}{Composite-Code} & CONCODE-GPT & 0.96 & 0.22 & 0.05 & \textbf{0.99} & 0.76\tabularnewline
 & Doc2Code-GPT & 0.94 & 0.34 & 0.04 & \textbf{0.99} & 0.97\tabularnewline
 & APPS-GPT & 0.76 & 0.64 & 0.07 & 0.98 & 0.86\tabularnewline
\hline 
\end{tabular}

}
\end{table}

\subsection{RQ4: Human Study}

\begin{figure}[!t]
    \centering
    \includegraphics[scale=0.42]{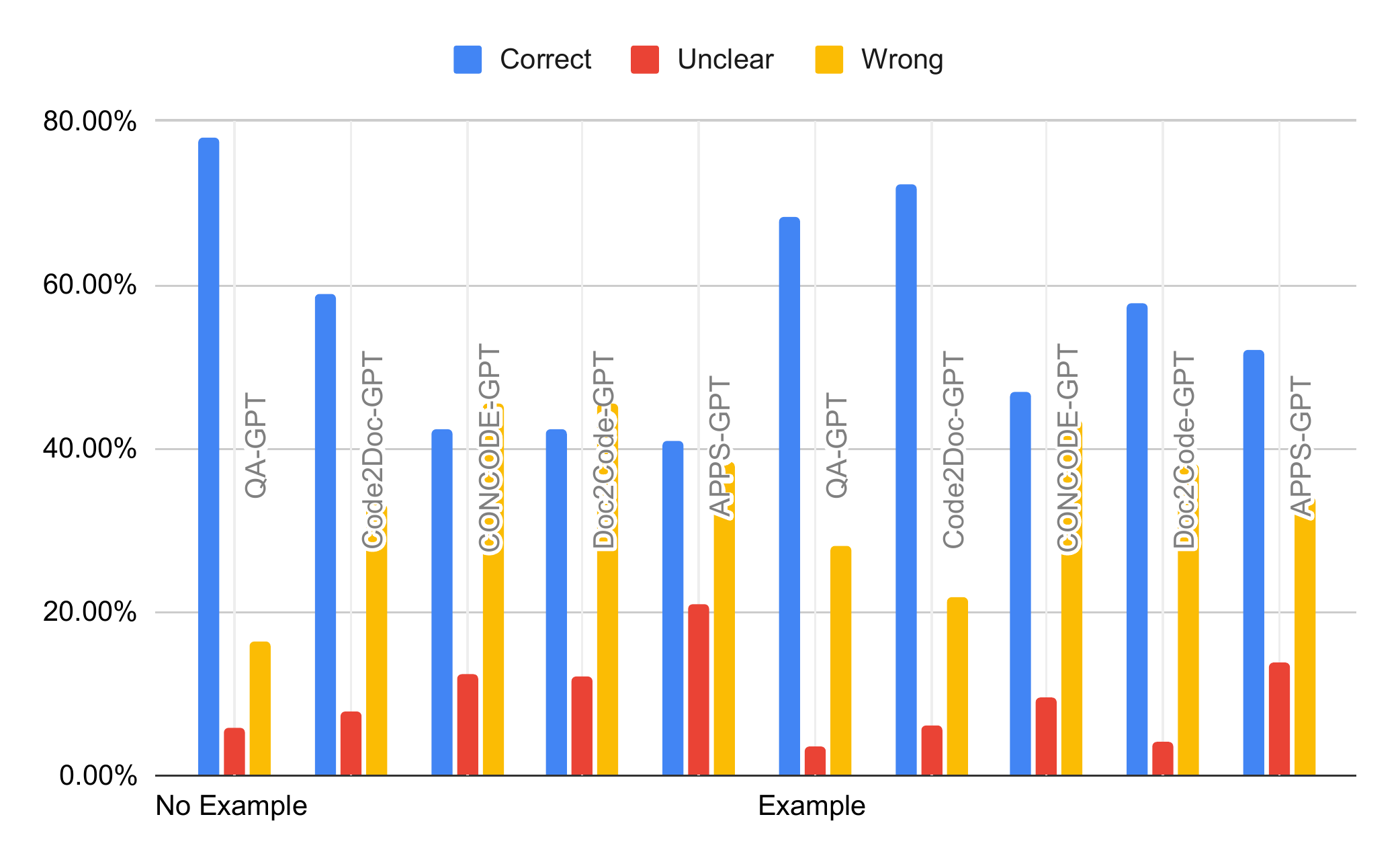}
    \caption{Performance of human participants in the survey.}
    \label{fig:hum-study}
\end{figure}

The results collected from our online survey are presented in Figure~\ref{fig:hum-study}.
We show the average accuracy of both the ``Example'' and ``No-example'' groups on each type of code-related task.
Overall, the ``Example'' group demonstrated a slightly better accuracy of 59.5\% vs. 52.5\% obtained by the ``No-Example'' group, across all tasks.
Both groups did well on Q\&A-GPT and Code2Doc-GPT, with an accuracy close to or above 60\% and as high as 77.9\%.
The contents presented to the participants in those two tasks only contain natural language texts
and do not contain any code snippets.
An explanation for the better performance could be that natural language texts may reveal more
hints in terms of the language patterns used, tone, and emotion conveyed.
The survey respondents also ranked ``language use---repetitive/formulaic language patterns'' as the most important factor making them believe a piece of content is generated by AI, among others, including ``coherence and structure'', ``tone and voice'' and ``emotional appeal''.
For tasks involving only code, the respondents did not perform as well.
On average, both groups got 47.1\% of the questions correct, which is similar to random guesses.

We also compared the performance of human with some best-performing detectors on the same set of questions. 
On an interesting note, we also tested the performance of ChatGPT in identifying the contents generated by itself or human, by asking it the survey questions.
\Cref{tbl-human-results} shows a summary of the comparison results. The accuracy is shown as the percentage of the correct answers to the survey questions.
For natural-language contents, our human respondents performed comparably well with most detectors (except Comp-NL which was especially fine-tuned on the same dateset).
For code snippets, the human subjects struggled and failed behind AITextClassifier, one of the best commercial detectors. The performance of ChatGPT was not ideal in this experiment.

\begin{table}[!t]
\caption{Performance comparison between human and detectors on the survey questions.}
\label{tbl-human-results}

\resizebox{1\columnwidth}{!}{

\begin{tabular}{l|ccccc}
\hline 
Scenarios & AITextClassifier & ChatGPT & Human & Comp-Code & Comp-NL\\
\hline 
\hline 
Q\&A-GPT & 0.60 & 0.38 & 0.68 & 0.70 & \textbf{1.00}\tabularnewline
Code2Doc-GPT & 0.50 & 0.42 & \textbf{0.72} & 0.50 & 1.00\tabularnewline
\hline 
CONCODE-GPT & \textbf{0.70} & 0.50 & 0.47 & \textbf{1.00} & 0.30\tabularnewline
Doc2Code-GPT & 0.40 & 0.44 & 0.58 & 0.90 & 0.40\tabularnewline
APPS-GPT & 0.60 & \textbf{0.64} & 0.52 & 0.60 & 0.40\tabularnewline
\hline 
\end{tabular}

}
\end{table}

 \vspace{5pt}\noindent \fbox{
	\parbox{0.95\linewidth}{\textbf{Answers to RQ4}: Similar to the findings of existing detectors, humans are better at distinguishing natural language data than code data. Some commercial detectors even have better performance than humans in detecting ChatGPT-generated code.}
}

%% file: Sections/discussion.tex

\subsection{Internal Validity}
First, the prompts we used to generate the ChatGPT contents may affect our results. We designed our prompts to mimic what an average user may provide to ChatGPT under the corresponding usage scenarios.
These may not always be the most representative ones. We plan to investigate the effects of different prompts in the future. 
Furthermore, the answers generated by ChatGPT are non-deterministic. 
Different answers may be generated even for the same prompt. 
But for the purpose of constructing datasets to evaluate AIGC detectors, the impact of non-determinism is limited. 
Second, we could not always verify the source of human-provided data in our dataset.
For instance, we considered the answers in the Stack Overflow dataset to be provided by human users, but this may not always be true. Some answers could be generated by other tools.
Similarly, treating the code snippets in APPS, CONCODE, and Code2Doc as human-written may not always be reliable.
But since these datasets were mostly populated with data generated before LLMs became mainstream, we estimate the impact to be limited. Finally, the detectors we studied require detection thresholds to be set and any threshold chosen may not always work the best in different settings. We acknowledge this potential threat and used the recommended values for each detector to mitigate this issue.

\subsection{External Validity}
First, the results obtained on the datasets we used and the detectors we studied may not be generalizable to other data and tools. To mitigate this threat, we collected data from different software development scenarios and contexts to make it more representative of the real-world development practice.
We selected both commercial and open-source detectors, which we believe represent the state of the art. Second, the recruitment of participants in our human study may have selection bias, thus affecting the external validity.
Since our survey was conducted online and we did not have total control on the population of the respondents.
To mitigate this, we ensured that the survey was distributed widely, to people who meet our selection criteria, namely, software developers/researchers having at least five years of programming experiences.

%% file: Sections/related.tex

\subsection{AI in Software Engineering}
In the early stages of this field, some smaller neural networks with fewer parameters were used to solve the problems in software engineering such as source code summarization~\cite{liu2021retrievalaugmented,ahmad2020transformer}, vulnerability detection~\cite{zhou2019devign, li2018vuldeepecker} and code search~\cite{gu2018deep, liu2023graphsearchnet}. Different neural network architectures are used such as LSTMs~\cite{hochreiter1997long}, Transformer~\cite{vaswani2017attention}, Graph Neural Networks~\cite{li2015gated}. With the development of this field, some more advanced techniques are adopted to achieve higher performance such as pre-training. The early works for code pre-trained models such as CodeBERT~\cite{feng2020codebert}, GraphCodeBERT~\cite{guo2020graphcodebert} takes encoder-only Transformer as the architecture to pre-train a general model on the code-related data and then fine-tune this pre-trained model to downstream tasks to achieve superior performance. The subsequent work has made further improvements such as PLBART~\cite{ahmad2021unified} and CodeT5~\cite{wang2021codet5} which use encoder-decoder Transformer as the model architecture to improve the model capacity. Although these pre-trained models have shown significant improvements in different software engineering tasks compared with previous works, they still cannot be applied in a real scenario. OpenAI takes a further step and released CodeX model~\cite{chen2022codet}, which is trained from GPT model on publicly available code from GitHub. Furthermore, a distinct production version of CodeX powers GitHub Copilot. ChatGPT released by OpenAI is another representative code, which is fine-tuned from GPT-3.5 series with RLHF for the alignment. The massive learnt knowledge in GPT-3.5 series and the powerful conversation ability enable ChatGPT to generate accurate answers in different domains. Hence, it can also be applied to software engineering such as code summarization as
well as generation~\cite{nair2023generating}, and vulnerability detection as well as repair~\cite{sobania2023analysis, surameery2023use}. For example, Sobania et al.~\cite{sobania2023analysis} utilized ChatGPT to fix bugs on the standard bug-fixing benchmark QuixBugs~\cite{lin2017quixbugs} and outperformed the state of the art, managing to fix 31 out of 40 bugs. It is precise because ChatGPT has been widely used in software engineering, in this work, we explore whether the ChatGPT-generated code can be detected by existing detectors.

\subsection{DeepFake Detection}
Deepfake refers to the creation or manipulation of facial appearance through deep generative approaches and deepfake detection aims to identify whether an image or video is synthesized with AI or produced naturally with a camera, which is similar to AIGC content detector. Based on the extracted features, they can be mainly categorized into spatial-based, frequency-based and biological signal-based. Detecting deepfake on the spatial domain is the most popular technique in the existing studies~\cite{li2020identification, wang2020cnn} and it aims to observe various visible or invisible artifacts on the spatial domain for distinguishing real and fake. Apart from the spatial domain, because the differences between real and synthesized fake faces can also be revealed in the frequency domain, there are also some studies~\cite{qian2020thinking, frank2020leveraging} exploiting the differences from the frequency domain. Furthermore, as real facial images and videos produced with cameras are natural compared to synthesized fake faces, hence biological signals can be used for distinguishing~\cite{li2018exposing, ciftci2020fakecatcher}. Compared with deepfake, where the detected objects are images or videos, we aim to identify the text content synthesized by ChatGPT.

%% file: Sections/conclusion.tex
In this paper, we present the first empirical study evaluating the performance of existing AIGC detectors in the software domain. We create a comprehensive dataset of code-related content generated by ChatGPT. The results of the study indicate that current AIGC detectors struggle with code-related data compared to natural language data. While fine-tuning can improve performance, the generalization of the model still remains a challenge. Furthermore, a human study is conducted to understand human capability in detecting ChatGPT-generated content, which indicates that humans also face similar challenges. The findings highlight the need for further research in this area, specifically the development of robust and generalized AIGC detectors.